\documentclass[twocolumn]{svjour3}          
\smartqed  
\usepackage{graphicx}
\begin{document}
\title{A discontinuity in the electromagnetic field of a uniformly accelerated charge}
\titlerunning{Discontinuity in the field of a uniformly accelerated charge}  
\author{Ashok K. Singal}
\authorrunning{A. K. Singal} 
\institute{A. K. Singal \at
Astronomy and Astrophysics Division, Physical Research Laboratory,
Navrangpura, Ahmedabad - 380 009, India\\
\email{ashokkumar.singal@gmail.com}         }
\date{Received: date / Accepted: date}
\maketitle
\begin{abstract}
The electric field of a uniformly accelerated charge shows a plane of discontinuity, where the field extending only on one side of the plane, terminates abruptly on the plane with a finite value. This indicates a non-zero divergence of the electric field in a source-free region, implying a violation of Gauss law.
In order to make the field compliant with Maxwell's equations everywhere, an additional field component, proportional to a $\delta$-function at the plane of discontinuity, is required.  Such a ``$\delta$-field'' might be the electromagnetic field of the charge, moving with a uniform velocity approaching $c$, the speed of light, prior to the imposition of acceleration at infinity. However, a range of attempts to derive this $\delta$-field for such a case, have not been entirely successful. Some of the claims of the derivation involve elaborate calculations with some not-so-obvious mathematical approximations. Since the result to be derived is already known from the constraint of its compliance with Maxwell's equations, and the derivation involves the familiar text-book expressions for the field of a uniformly moving charge, one would expect an easy, simple approach, to lead to the correct result. Here, starting from the electromagnetic field of a uniformly accelerated charge in the instantaneous rest frame, in terms of the position and motion of the charge at the retarded time, we derive this $\delta$-field, consistent with Maxwell's equations, in a fairly simple manner. This is followed by  a calculation of the energy in the $\delta$-field, in an analytical manner without making any approximation, where we show that this energy is exactly the one that would be lost by the charge because of the radiation reaction on the charge, proportional to its rate of change of acceleration, that was imposed on it at a distant past.
\end{abstract}
\section{Introduction}
Electromagnetic field of a charge undergoing a uniform proper acceleration was derived first by Max Born \cite {32} in 1909, and later by many more authors \cite{SO10,Lau11,7}. From  Born's solution, Pauli \cite{33} in 1921 noticed  that at a specific instant, when the charge was instantly stationary,  the magnetic field was zero everywhere and from that he concluded that there can be no radiation from a uniformly accelerated charge. 

Bondi and Gold \cite{3} in 1955 first drew attention to an anomaly apparent in Born's solution that the electric field component along the direction of motion terminates with a non-zero value at a plane normal to the direction of motion; the discontinuity of electric field implying a breakdown of one of Maxwell's equations, viz. Gauss law. Such a discontinuity, would normally be accompanying a surface charge density at the plane of discontinuity \cite{19}. 
However, as no such charge density is pre-specified in the present case, one could be led to the conclusion that the Maxwell's equations are incompatible with the existence of a {\em single} charge, uniformly accelerated at all times.

Bondi and Gold \cite{3} proposed that a consistency with Maxwell's equations could be restored if owing to some process there were a certain $\delta$-function field at the plane of discontinuity. One scenario suggested for this purpose was that prior to some instant in distant past, say $t=-\tau$, the charge moved with a uniform velocity and after $t=-\tau$, with a constant acceleration. Then in the limit $\tau \rightarrow \infty$, one may obtain the required ``$\delta$-field.'' Though they themselves \cite{3} did not further pursue this line of thinking, but afterwards in literature this seems to have become the standard explanation \cite{5,10,Ly08,ER00} for the $\delta$-field, required to make total field conform to Gauss law. 

However, one of the exasperating aspect of this is that a simple derivation of the $\delta$-field from the electromagnetic field of a charge that moved with a uniform velocity before $t=-\tau$ with ${\tau \rightarrow \infty}$ in the limit, has not been always successful. 
Since the result to be obtained is already known from the constraint of its compliance with Maxwell's equations, and the derivation involves the familiar, text-book expressions for the field of a uniformly moving charge, one would expect a simple, an almost ingenuous approach, to lead to a correct result. This two stage motion has been termed `truncated hyperbolic motion'  and a recent attempt to derive the $\delta$-field from the truncated motion, using a simple method, remained  unsuccessful \cite{FR14,FR15}. The $\delta$-field was derived by Cross \cite{CR15}, but the derivation involved a rather complicated-looking evaluation of asymptotic terms in the Li\'enard-Wiechert potentials. Continuous electric field lines for
truncated motion were found recently, and the electric field $\delta$-function derived in the limit $\tau\rightarrow\infty$ \cite{JF15}. Our derivation here arrives at similar results by a different method where we shall compare the two field expressions in terms of motion at retarded time, and show that way why the two fields differ in their transverse components; in the case of the uniformly accelerated charge, the transverse component is zero as the acceleration field gets neatly cancelled by the transverse component of the velocity field in the instantaneous rest frame of the charge.

We will use here the truncated hyperbolic motion to derive the $\delta$-field. For this, we shall first derive the electromagnetic field of a uniformly accelerated charge in the instantaneous rest frame, in terms of the position and motion of the charge at the retarded time. From that, using an algebraic transformation, the field in terms of ``real-time'' position of the charge will be arrived at. This procedure allows us a better understanding of the interrelation between the field of a uniformly accelerated charge and that of a uniformly moving charge, especially in the region of their common boundary. We may state that a computation of the total electromagnetic energy in the field of a uniformly accelerated charge and its comparison with the total energy in the  electromagnetic field of a charge moving with a uniform velocity, but with a magnitude equal to the ``present speed'' of the  uniformly accelerated charge, has shown that  in both cases the two quantities are exactly the same \cite{17,18}. Similar is the conclusion arrived at for the net Poynting flux through a given spherical surface centered on the retarded time position of the charge, from a comparison between the two cases \cite{18}.

Here we will not only derive the $\delta$-field, arising from the uniform motion of the charge before $t=-\tau$, with ${v \rightarrow c}$ for ${\tau \rightarrow \infty}$, but shall follow it with a computation of the total energy in the field, in an exact analytical manner, showing that the energy in $\delta$-field is the one lost by the moving charge due to radiation reaction at time $t=-\tau$ in the limit ${\tau \rightarrow \infty}$, when there was a rate of change of acceleration of the charge.
\section{Electromagnetic field of a charge undergoing a uniform proper acceleration}
From a uniformly accelerated motion we mean a motion with a constant proper acceleration. We may  take it to be a one-dimensional motion since an appropriate Lorentz transformation can always be employed  so as to reduce the velocity component normal to the acceleration vector to zero. 

The electromagnetic field of a charge $e$, moving with ${\bf v}\parallel \dot{\bf v}$, can be written for any given time $t$ as \cite{1,2,25,28}
\begin{eqnarray}
\label{eq:38aa}
{\bf E}=\left[\frac{e({\bf n}-{\bf v}/c)}
{\gamma ^{2}r^{2}(1-{\bf n}\cdot{\bf v}/c)^{3}} +\frac{e\:{\bf n}\times({\bf n}\times
\dot{\bf v})}{rc^2\:(1-{\bf n}\cdot
{\bf v}/c)^{3}}\right]_{t_{\rm r}}\;,
\end{eqnarray}
where quantities within the square brackets, unless otherwise specified, are to be evaluated at the corresponding retarded time $t_{\rm r}=t-r/c$. The acceleration field (the second term within the square brackets),  transverse to ${\bf n}$ and falling with distance as $1/r$, is usually assumed to be responsible for radiation from a charge. 

Using the vector identity ${\bf v}={\bf n}({\bf v}.{\bf n}) - {\bf n}\times\{{\bf n}\times{\bf v}\}$, we can rewrite the velocity field (first term within the square brackets), in terms of  radial (along {\bf n}) and transverse components (normal to {\bf n}) as \cite{17,18}
\begin{eqnarray}
\label{eq:38bb}
\left[\frac{e({\bf n}-{\bf v}/c)}{\gamma ^{2}r^{2}(1-{\bf n}\cdot{\bf v}/c)^{3}}\right]_{t_{\rm r}}
&=&\left[\frac{e\:{\bf n}}{\gamma ^2 r^2(1-{\bf n}\cdot{\bf v}/c)^2}\right.\nonumber\\
&&\left.+\frac{e\:{\bf n}\times({\bf n}\times\gamma{\bf v})}
{\gamma^3r^2c\:(1-{\bf n}\cdot{\bf v}/c)^3}\right]_{t_{\rm r}}.
\end{eqnarray}

This is also the expression for the electric field of a charge moving with a uniform velocity $\bf v$. Thus the total electric field of a uniformly moving charge or the velocity field part of an accelerated charge, both have radial as well as transverse components, with respect to the time-retarded charge positions.

Now, for an assumed one dimensional motion with a uniform proper acceleration, ${\bf a} =\gamma^{3} \dot{\bf v}$, the `` present'' velocity ${{\bf v}}_{\rm o}$ at time $t$ is related to its value ${{\bf v}_{\rm r}}$ at the retarded time $t_{\rm r}=t-r/c$ as 
\begin{eqnarray}
\label{eq:38c}
\gamma_0 {\bf v}_0=\gamma_{\rm r} {\bf v}_{\rm r}+ (t-t_{\rm r}){\bf a} =\left[\gamma {\bf v}+\frac{r \gamma^{3} \dot{\bf v}}{c}\right]_{t_{\rm r}}\:,
\end{eqnarray}
where $\gamma_0$, $\gamma_{\rm r}$ are the corresponding Lorentz factors. 

Substituting Eq.~(\ref{eq:38bb}) in Eq.~(\ref{eq:38aa}), and utilizing Eq.~(\ref{eq:38c}), we find that the net electric field of a uniformly accelerated charge is 
\begin{eqnarray}
\label{eq:38aaa}
{\bf E}=\left[\frac{e{\bf n}}{\gamma ^2 r^2(1-{\bf n}\cdot{\bf v}/c)^2}+\frac{e{\bf n}\times({\bf n}\times \gamma_0{\bf v}_0/c)}{\gamma ^{3}r^{2}(1-{\bf n}\cdot{\bf v}/c)^{3}} \right]_{t_{\rm r}}\:.
\end{eqnarray}
It should be noted that $\gamma_0$ and ${\bf v}_0$ are the values at `present' time $t$.

The magnetic field can be computed from the electric field \cite{1,2,25,28}, which in the case of a uniformly accelerated charge, turns out from Eq.~(\ref{eq:38aaa}), to be 
\begin{eqnarray}
\label{eq:38ba}
{\bf B}=[{\bf n}] \times {\bf E}=\left[\frac{-e{\bf n}\times \gamma_0{\bf v}_0/c}{\gamma ^{3}r^{2}(1-{\bf n}\cdot{\bf v}/c)^{3}} \right]_{t_{\rm r}}\:.
\end{eqnarray}

In the instantaneous rest-frame, where, by definition, the present velocity ${{\bf v}_0}=0$, we have then a simple expression for the total electric field of a uniformly accelerated charge as
\begin{eqnarray}
\label{eq:38ab}
{\bf E}=\left[\frac{e{\bf n}}{\gamma ^2 r^2(1-{\bf n}\cdot{\bf v}/c)^2}\right]_{t_{\rm r}},
\end{eqnarray}
where there is only a radial electric field with respect to the charge position at the  retarded time.
There is of course, a nil magnetic field, ${\bf B}=0$, in the instantaneous rest-frame.
\section{Electromagnetic field in the instantaneous rest frame of a uniformly accelerated charge in terms of its ``present'' position}
As is well known, the only non-trivial case of uniform acceleration 
is that of a hyperbolic motion \cite{MTW73,11a,GO82}. 
We assume that $a\equiv \gamma ^{3} \dot{v}$ is the uniform acceleration along the z-axis. 
Let the charge initially moving along the $-z$ direction from $z=\infty$ at time $t=-\infty$, is getting constantly decelerated till, say, at time $t=0$ it comes to rest momentarily at $z=z_0$, and then onwards moves with an increasing velocity along the $+z$ direction.
The motion of the charge for $t>0$ follows exactly that of a relativistic rocket with a uniform proper acceleration $a$, starting from $z_0$ at $t=0$ \cite{MTW73,11a,GO82,DS98,83}.
Due to the cylindrical symmetry of the system, it is convenient to employ  cylindrical coordinates ($\rho,\phi,z$). Without any loss of generality, we can choose the origin of the coordinate system, $z=0$, at $z_0-c^{2}/a$.

Let $z_{\rm t}$ be the position of the charge at a time $t$, where it has the velocity $v_{\rm t} = {a}t/\gamma_{\rm t}$,  with 
\begin{eqnarray}
\label{eq:38ac}
\gamma_{\rm t} =\left[{1+(at/c)^2}\right]^{1/2} =\left[{1+(ct/z_0)^2}\right]^{1/2}. 
\end{eqnarray}
Then we have
\begin{eqnarray}
\label{eq:38ad}
z_{\rm t}&=&z_0 + \int_{0}^{t} \frac{at}{\left[{1+({a}t/c)^2}\right]^{1/2}} \:{\rm d}t\nonumber\\
&=&z_0+ \frac{c^2}{a} \left[\left\{{1+({a}t/c)^2}\right\}^{1/2} -1\right]\nonumber\\
&=&\left[{z_0^2+c^2t^2}\right]^{1/2}=z_0\gamma_{\rm t}\,.
\end{eqnarray}
The choice of origin, so that $z_0=c^2/a$, makes the relation between charge position and time a rather convenient one.  

The equation~(\ref{eq:38ad}) , written as $z_{\rm t}^2- c^2t^2= z_0^2$, can be recognized as the equation of a hyperbola in space-time ($z_{\rm t},t$) coordinates, with $z_{\rm t}=z_0$ at $t=0$.   

We can express the electromagnetic field of a uniformly accelerated charge, not necessarily in terms of motion of the charge at retarded time as in Eqs.~(\ref{eq:38aaa}), (\ref{eq:38ba}) or (\ref{eq:38ab}), instead in terms of the ``real-time'' value of the charge position. 
Here we shall, starting from Eq.~(\ref{eq:38ab}), determine the electromagnetic field in the instantaneous rest frame of the uniformly accelerated charge, where $v_0=0$ at $t=0$. 

If $P$ ($\rho,z$) be the field point, where we wish to determine the electromagnetic field at a time $t=0$, in terms of the ``real-time'' position of the charge $z_0$ when its velocity $v_0=0$, all we require is the quantity $r\gamma_{\rm r}(1-v_{\rm r} \cos\theta/c)$ in  Eq.~(\ref{eq:38ab}) to be expressed in terms of $\rho$, $z$ and $z_0$. (For the field expression in a more general case, i.e. at $t\ne 0$ when the uniformly accelerated charge may be moving with some finite velocity, we begin with  Eqs.~(\ref{eq:38aaa}) and (\ref{eq:38ba}), see the Appendix.)  

\begin{figure*}[t]
\begin{center}
\includegraphics[width=\linewidth]{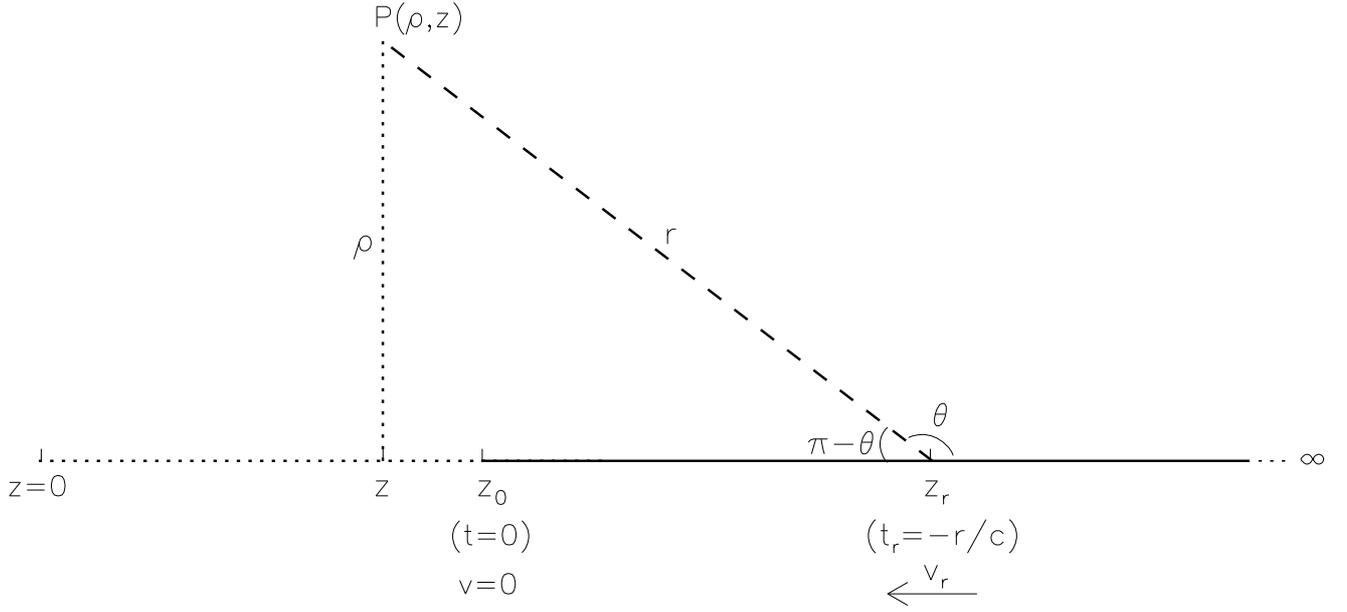}
\caption{A charge, moving with a uniform proper acceleration, ${\bf a} =\gamma^{3} \dot{\bf v}$, along the z-axis, and coming from $z=\infty$ at $t=-\infty$ with an initial velocity along  $-z$ direction, comes to rest momentarily at $z=z_0$ at time  $t=0$. We want to determine the  electromagnetic field at time $t=0$, at a point $P (\rho,z)$, which is at a distance $r$ from the charge position $z_{\rm r}$ at the corresponding retarded time $t_{\rm r}=-r/c$, when it was moving with a velocity $v_{\rm r}$ along $-z$ direction.}
\end{center}
\end{figure*}

Let $z_{\rm r}=z_0\gamma_{\rm r}$ be the position of the charge at the corresponding retarded time $t_{\rm r}=-r/c$, where it had the velocity 
\begin{eqnarray}
\label{eq:38.0}
v_{\rm r} = -\frac{a r}{c\gamma_{\rm r}} = -\frac{c r}{z_0\gamma_{\rm r}}
\end{eqnarray}
the negative sign here indicates the velocity is along $-z$ direction (Fig.~1). Then we have
\begin{eqnarray}
\label{eq:38}
z_{\rm r}^2=z_0^2\gamma_{\rm r}^2=z_0^2 \left[{1+\left(\frac{ct_{\rm r}}{z_0}\right)^2} \right]=z_0^2+r^2.
\end{eqnarray}
From Fig. 1, we have the relations
\begin{eqnarray}
\label{eq:38.1}
z_{\rm r}=z-r\cos\theta, \,\,\,\,\,\, \rho=r\sin\theta\,,
\end{eqnarray}
where $\theta$ is the angle with respect to the z-axis. 
This gives us
\begin{eqnarray}
\label{eq:38a}
z_{\rm r}^2+\rho^2=z^2+r^2-2 r z \cos \theta\,.
\end{eqnarray}
From Eqs.~(\ref{eq:38}) and (\ref{eq:38a}), we get
\begin{eqnarray}
\label{eq:38a1}
2rz\cos \theta=-(z_0^{2}+\rho^{2}-z^{2})\,.
\end{eqnarray}
After squaring, and using Eq.~(\ref{eq:38.1}), we can obtain
\begin{eqnarray}
\label{eq:38a2}
r=\frac{[(z_0^{2}+\rho^{2}-z^{2})^2+4z^{2}\rho^{2}]^{1/2}}{2z}\,.
\end{eqnarray}
Also 
\begin{eqnarray}
\label{eq:38b2}
\gamma_{\rm r}(1-v_{\rm r} \cos\theta/c)
=\frac{z_{\rm r}}{z_0}+\frac{r\cos\theta}{z_0}=\frac{z}{z_0}\,.
\end{eqnarray}
From Eqs.~(\ref{eq:38a2}) and (\ref{eq:38b2}), we get
\begin{eqnarray}
\label{eq:38a4}
r\gamma_{\rm r}(1-v_{\rm r} \cos\theta/c)&=&\frac{[(z_0^{2}+\rho^{2}-z^{2})^2+4z^{2}\rho^{2}]^{1/2}}{2z_0}\nonumber\\
&=&\frac{\xi_0}{2z_0}\,,
\end{eqnarray}
where 
\begin{eqnarray}
\label{eq:38ab2}
\xi_0=[(z_0^{2}+\rho^{2}-z^{2})^2+4z^{2}\rho^{2}]^{1/2}
\,.
\end{eqnarray}

Substituting Eqs.~(\ref{eq:38a4}) in Eq.~(\ref{eq:38ab}), we get  the electromagnetic field at $P(\rho,z)$ at time $t=0$ as 
\begin{eqnarray}
\label{eq:38a5}
{\bf E}=\frac{4 e z_0^2 {\bf n}}{\xi_0^2}\,,
\end{eqnarray}

Using (\ref{eq:38a1}) and (\ref{eq:38a2}), the field components are then written as 
\begin{eqnarray}
\label{eq:38ab1}
\nonumber
E_{z}&=& \frac{4 e z_0^2\cos\theta}{\xi_0^2}
=\frac{-4ez_0^{2}(z_0^{2}+\rho^{2}-z^{2})}{\xi_0^{3}}\\
E_{\rho}&=&\frac{4 e z_0^2}{\xi_0^2}\frac{\rho}{r}
=\frac{8ez_0^{2}\rho z}{\xi_0^{3}}\,,
\end{eqnarray}
the remaining  field components being zero. 

The above field expressions are in the instantaneous rest frame of the uniformly accelerated charge, and agree with those derived earlier \cite{FR14,JF15}.
For the more general case, when the uniformly accelerated charge, at $t\ne 0$, would be located elsewhere and moving with a finite velocity, the derivation of field expressions, though on similar lines, is a bit more involved and is given in the Appendix. 

\section{A discontinuity in electromagnetic field of a uniformly accelerated charge at the $z=0$ plane}
From causality arguments, field of a uniformly accelerated charge, beginning from an infinite past, are to be found only in the region $z>0$ \cite{5}.
In the limit $z\rightarrow 0$, from Eq.~(\ref{eq:38ab1}), we can write the field as
\begin{eqnarray}
\label{eq:38.ab1}
E_{\rho}&=&0\nonumber\\
E_{z}&=& \frac{-4ez_0^2}{(z_0^{2}+\rho^{2})^2}\,.
\end{eqnarray}
\begin{figure*}[t]
\begin{center}
\includegraphics[width=\linewidth]{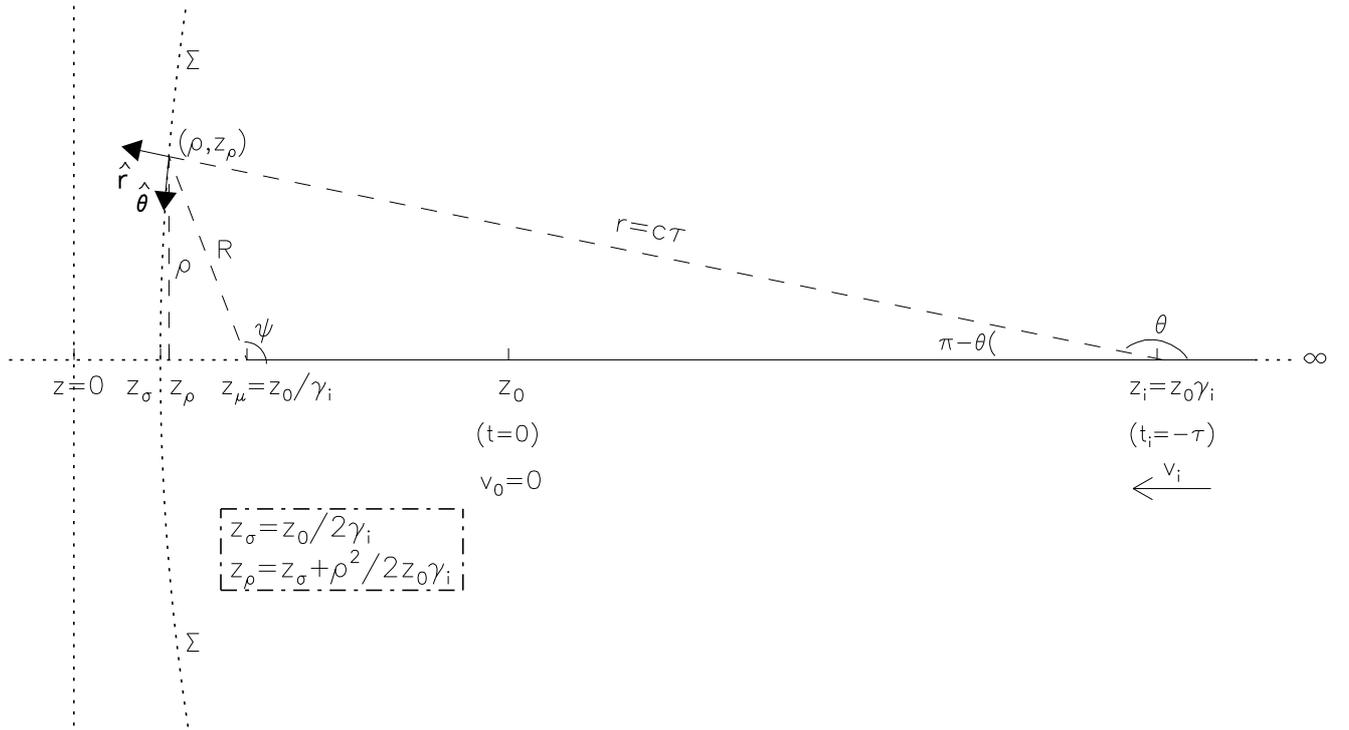}
\caption{A charge coming from infinity ($z=\infty$ at $t=-\infty$) with an initial uniform velocity $v_{\rm i}$, with a corresponding Lorentz factor $\gamma_{\rm i}$, moving along  $-z$ direction, is applied a constant proper acceleration, ${\bf a} =\gamma^{3} \dot{\bf v}$ along the z-axis, at time $t_{\rm i}=-\tau$ when it is at a location $z_{\rm i}$, and thereafter comes to rest ($v_0=0$) momentarily, at $z=z_0$ at time  $t=0$. 
The charge would have reached its extrapolated ``present'' position $z_\mu$ at time  $t=0$, if it had continued to move with its initial uniform velocity $v_{\rm i}$. 
At time $t=0$, the electromagnetic field at points just within $\Sigma$, the boundary of a  sphere of radius $r=c \tau=z_0\gamma_{\rm i} |v_{\rm i}|/c$, centered on the charge position $z_{\rm i}$,  is that of a uniformly accelerated charge, with only a radial component along $\bf \hat{r}$. However, at points just outside the sphere $\Sigma$, the field is along $\bf \hat{R}$, in radial direction from the ``present'' charge position $z_\mu$, but having both a radial component along $\bf \hat{r}$ and a transverse component along $\hat{{\mbox{\boldmath $\theta$}}}$ with respect to the corresponding retarded charge position $z_{\rm i}$.
The spherical boundary $\Sigma$ intersects the z-axis at a point $z_\sigma$. The box shows $z_\sigma$ and $z_\rho$ both expanded upto first order terms in $1/\gamma_{\rm i}$. The plot in the figure is for $\gamma_{\rm i}=2.5$, corresponding to a velocity $|v_{\rm i}|\approx 0.92 c$.}
\end{center}
\end{figure*}

Since the field of the uniformly accelerated charge at $t=0$ does not extend into the region $z\le 0$, the field component $E_{z}$ in Eq.~(\ref{eq:38.ab1}) terminates with a finite value as the $z=0$ plane is approached and there is thus a discontinuity in electromagnetic field of a uniformly accelerated charge at the $z=0$ plane. Of course this could happen if there were a surface charge density
\begin{eqnarray}
\label{eq:88.4b}
\sigma=\frac{1}{4\pi }\left.{E_z} \right|_{\rm z\rightarrow 0} =\frac{-e z_0^{2}}{\pi(z_0^{2}+\rho^{2})^2}\,
\end{eqnarray}
at the $z=0$ plane, amounting to a total charge $-e$. There is of course no surface charge density at the $z=0$ plane pre-assigned in our case, where the only charge in the picture is the one with uniform acceleration. Therefore this discontinuity at the $z=0$ plane implies a violation of one of Maxwell's equations, viz. Gauss law.

However, compatibility with the Maxwell's equations could be restored if at the $z=0$ plane, there were additional field proportional to the  $\delta$-function \cite{3}
\begin{eqnarray}
\label{eq:88.4a}
E_{\rho}  =  \frac{2e \rho}{z_0^{2}+\rho^{2}} \delta(z) 
\;.
\end{eqnarray}

To verify, the contribution of the ``$\delta$-field'' (Eq.~(\ref{eq:88.4a})) to $\nabla \cdot {\bf E}$ at the $z=0$ plane is
\begin{eqnarray}
\label{eq:38.4c}
\left.\frac{1}{\rho}\frac{\partial (\rho E_{\rho})} {\partial \rho}\right|_{\rm z=0}=\frac{4e z_0^{2}}{(z_0^{2}+\rho^{2})^2}\delta(z)\,
\end{eqnarray}
which neatly cancels the surface charge density inferred in (Eq.~(\ref{eq:88.4b})), and we then obtain $\nabla \cdot {\bf E}= 0$ at the $z=0$ plane. 

We want to derive $\delta$-field (Eq.~(\ref{eq:88.4a})) from the charge that moved with a uniform velocity prior to some instant in distant past, and afterwards has been moving with a constant acceleration. Let us assume that the uniform acceleration was imposed
upon the charge starting from a time $t_{\rm i}=-\tau$ onwards and that until then the
charge, in continuation with the ensuing accelerated motion specifications, was moving with a constant velocity $v_{\rm i}=-c^2\tau/(z_0^{2}+{c^{2} \tau^{2}})^{1/2}$, with a corresponding Lorentz factor $\gamma_{\rm i}=[1+(c \tau/z_0)^2]^{1/2}$ (Fig.~2).
In that case, at time $t=0$ while field in the region $r<c\tau$
(i.e., in the region inside of the spherical surface $\Sigma$ of radius $r$ centred at $z_{\rm i}=
(z_0^{2}+{c^{2} \tau^{2}})^{1/2}$, position of the charge at time $t_{\rm i}=-\tau$)
is that of a uniformly accelerated charge, the field in the region
$r>c\tau=z_0\gamma_{\rm i}|v_{\rm i}|/c$ is that of a uniformly moving charge.
\begin{figure*}[t]
\begin{center}
\includegraphics[width=\linewidth]{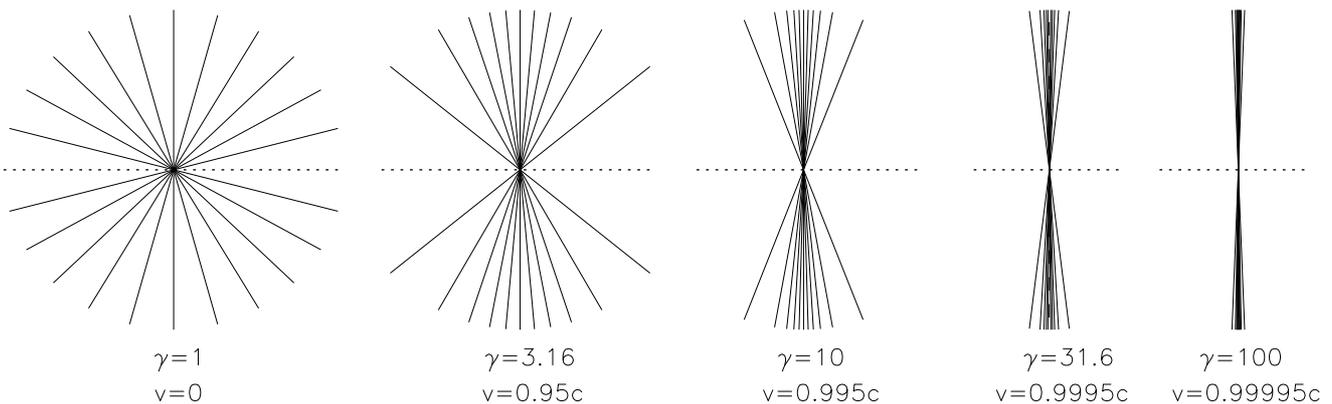}
\caption{The distribution of electric field lines of a charge moving with a uniform velocity $v$ and a corresponding  Lorentz factor $\gamma$. The field line distributions, centered on the ``present'' position of the moving charge, are shown for different Lorentz factors, with $\gamma$ increasing in steps of 3.16 $(=\sqrt{10})$. As $\gamma\rightarrow \infty$, with $v\rightarrow c$, the electric field increasingly resembles a $\delta$-field.}
\end{center}
\end{figure*}

The electric field just inside the boundary $\Sigma$, being that of the charge moving with a uniformly acceleration, from Eq.~(\ref{eq:38ab}), can be written as  
\begin{eqnarray}
\label{eq:38.5a}
{\bf E}_{\rm a}=\left[\frac{e{\bf \hat{r}}}{\gamma_{\rm i} ^2 r^2(1-{\bf n}\cdot{\bf v}_{\rm i}/c)^2}\right]_{t_{\rm i}},
\end{eqnarray}
while, the electric field just outside the boundary $\Sigma$, being that of the charge moving with a uniform velocity $v_{\rm i}$, from Eq.~(\ref{eq:38bb}) can be written as
\begin{eqnarray}
\label{eq:38.5b}
{\bf E}_{\rm v}&=&\left[\frac{e\:{\bf\hat{r}}}{\gamma_{\rm i} ^2 r^2(1-{\bf n}\cdot{\bf v}_{\rm i}/c)^2}\right]_{t_{\rm i}}\nonumber\\
&&+\left[\frac{e{v}_{\rm i}\sin \theta\:\hat{{\mbox{\boldmath $\theta$}}}}
{\gamma_{\rm i}^2r^2c\:(1-{\bf n}\cdot{\bf v}_{\rm i}/c)^3}\right]_{t_{\rm i}}.
\end{eqnarray}

The radial components (along ${\bf \hat{r}=n}$, Fig.~2) on the boundary $\Sigma$ are exactly the same in Eqs.~(\ref{eq:38.5a}) and (\ref{eq:38.5b}), implying a continuity in the radial components. We see that for a finite $\tau$, there is no discontinuity in the radial component across the spherical boundary $\Sigma$, even though the electric field of the uniformly accelerated charge terminates with a finite value at the boundary $\Sigma$. The difference in the electric fields at the boundary $\Sigma$ lies only in the transverse components; in the case of the uniformly moving  charge, there is a finite transverse component ($\propto {v}_{\rm i} \sin \theta$) along $\hat{{\mbox{\boldmath $\theta$}}}$ (Fig.~2), while in the case of the uniformly accelerated charge, the transverse component is zero as the acceleration field, responsible for radiation in a usual case, gets neatly cancelled by the transverse component of the velocity field at $t=0$, when the instantaneous velocity of the uniformly accelerated charge becomes zero.

The uniformly moving  charge has also a magnetic field $\propto {v}_{\rm i}\sin \theta$,
\begin{eqnarray}
\label{eq:38.5c}
{\bf B}=\left[\frac{e {v}_{\rm i}\sin \theta\:\bf \hat{{\mbox{\boldmath $\phi$}}}}{\gamma_{\rm i} ^{2}r^{2}c(1-{\bf n}\cdot{\bf v_{\rm i}}/c)^{3}} \right]_{t_{\rm i}}\:.
\end{eqnarray}
It should be noted that in Eqs.~(\ref{eq:38.5b}) and (\ref{eq:38.5c}), $v_{\rm i}=-c^2\tau/(z_0\gamma_{\rm i})$, is of a negative value.

The electric field due to the charge moving with uniform velocity ${\bf v}_{\rm i}$ before $t_{\rm i}=-\tau$, can be written equivalently, in an alternative form, with respect to the extrapolated ``present'' position $z_\mu$ of the charge. This is the position the charge would occupy if there were no acceleration imposed on it and it had continued to move with a constant velocity $v_{\rm i}$.  Then at $t=0$ we have, $z_\mu=z_{\rm i}+v_{\rm i}\tau =z_0 \gamma_{\rm i}-v_{\rm i}^2z_0\gamma_{\rm i}/c^2=z_0/\gamma_{\rm i}$, with 
the electric field of the charge being everywhere in a radial direction $\hat{\bf R}$ from $z_\mu$ \cite{1,2,25,PU85}.
\begin{eqnarray}
\label{eq:88.4d}
{\bf E}&=&\frac{e\hat{\bf R}}{R^{2}\gamma_{\rm i}^{2}[1-(v_{\rm i}/c)^{2}\sin^{2}\psi]^{3/2}}\nonumber\\
&=&\frac{e\gamma_{\rm i}{\bf R}}{[(\gamma_{\rm i}^2(z-z_\mu)^{2}+\rho^2]^{3/2}}\,,
\end{eqnarray}
with the magnetic field, ${\bf B}= {\bf v}_{\rm i}\times {\bf E}/c$ \cite{1,2,25}.

Figure~3 shows the distribution of electric field lines of a charge moving with a uniform velocity $v$ and a corresponding Lorentz factor $\gamma$. The field lines, centered on the ``present'' position of the moving charge, are shown for different velocities and the corresponding Lorentz factors. As $\gamma$ becomes larger, the electric field component along the direction of motion, becomes negligible relative to the perpendicular component, with field lines increasingly oriented perpendicular to the direction of motion. Moreover the field becomes negligible, except in a narrow zone along the direction of motion. In the limit $\gamma_{\rm i} \rightarrow \infty$, the electric field turns into `$\delta$-field' (Fig.~3), at the $z=0$ plane, representing the spherical surface, $\Sigma$ of infinite radius $r=c\tau$,  when $\tau \rightarrow \infty$. 

The question of discontinuity at the boundary $\Sigma$ arises only when $\tau\rightarrow\infty$, and as a result the spherical surface $\Sigma$ coincides with the $z=0$ plane. As $\gamma_{\rm i} \rightarrow \infty$ the field arising from the uniform motion of the charge before $t=-\tau$ become $\delta$-field, with a negligible $E_z$ component at $z=0$, and then the question of Gauss law violation at $z=0$ manifests itself.

It may be tempting to calculate the field by putting $z_\mu=z_0/\gamma_{\rm i}=0$ directly in Eq.~(\ref{eq:88.4d}), 
to get the $\delta$-field at the $z=0$ plane in the limit $\gamma_{\rm i}\rightarrow\infty$, however as has been shown \cite{FR14,FR15}, it does not lead to the correct expression for $\delta$-field this way. Instead, an appropriate limiting procedure must be followed to derive the $\delta$-field, as shown below. 
\subsection{A simple derivation of the $\delta$-field}
From Eq.~(\ref{eq:88.4d}), we can write the field components in terms of $\delta(z)$ as  
\begin{eqnarray}
\label{eq:a38.4e}
E_z=  \lim_{\gamma_{\rm i}\rightarrow\infty}\left[ \int_{-\infty}^{z_\rho} {\rm d}z\frac{e\gamma_{\rm i}(z-z_\mu)}{[(\gamma_{\rm i}^2(z-z_\mu)^{2}+\rho^2]^{3/2}}\right]\delta(z)\,,\nonumber \\
E_\rho= \lim_{\gamma_{\rm i}\rightarrow\infty}\left[\int_{-\infty}^{z_\rho} {\rm d}z\frac{e\gamma_{\rm i}\rho}{[(\gamma_{\rm i}^2(z-z_\mu)^{2}+\rho^2]^{3/2}}\right]\delta(z)\,.
\end{eqnarray}
Here limits of the integral are from $z=-\infty$ to $z_\rho$, where $z_\rho$ is the $z$ coordinate of the circle of radius $\rho$ on the spherical boundary $\Sigma$ (Fig.~2).

Substituting $z_\mu=z_0/\gamma_{\rm i}$ gives us the electric field components at time $t=0$ as
\begin{eqnarray}
\label{eq:a38.4f}
E_z&=&  \lim_{\gamma_{\rm i}\rightarrow\infty}\left[\frac{-e}{\gamma_{\rm i}[(\gamma_{\rm i}z_\rho-z_0)^{2}+\rho^2]^{1/2}}\right]\delta(z)\,,\nonumber \\
E_\rho&=& \frac{e}{\rho}\lim_{\gamma_{\rm i}\rightarrow\infty}\left[\frac{(\gamma_{\rm i}z_\rho-z_0)}{[(\gamma_{\rm i}z_\rho-z_0)^{2}+\rho^2]^{1/2}}+1\right]\delta(z)\,.
\end{eqnarray}

We want to determine the field components in the limit, $\gamma_{\rm i} \rightarrow \infty$,  
 with $z_{\rm i}=z_0\gamma_{\rm i} \rightarrow \infty$ and $z_\mu=z_0/\gamma_{\rm i}$ approaching the $z=0$ plane.

From Fig.~2, we have $(r-(z_\rho-z_\sigma))^2+\rho^2=r^2$, from which, for a fixed $\rho$, we get to a first order in $1/\gamma_{\rm i}$
\begin{eqnarray}
\label{eq:38.4g}
z_\rho-{z_\sigma}\approx\rho^2/2r\approx\rho^2/2\gamma_{\rm i}z_0\,.
\end{eqnarray}
Also, to a first order in $1/\gamma_{\rm i}$, we have
\begin{eqnarray}
\label{eq:38.4h}
z_\sigma=\gamma_{\rm i} z_0 (1-v_{\rm i}/c)=z_0/[\gamma_{\rm i}(1+v_{\rm i}/c)]\approx z_0/2\gamma_{\rm i}\,.
\end{eqnarray}
Thus we see that $z_\sigma\approx z_\mu/2$ and that all three $z_\sigma,z_\rho,z_\mu,$ always lie at $z>0$, however, in the limit $\gamma_{\rm i}\rightarrow 0$, they all approach the $z=0$ plane. 

We need to evaluate $\gamma_{\rm i}z_\rho-z_0$ in the limit $\gamma_{\rm i}\rightarrow\infty$.
From Eqs.~(\ref{eq:38.4g}) and (\ref{eq:38.4h}), where we kept terms upto first order in $1/\gamma_{\rm i}$, as those only would later survive when $\gamma_{\rm i}\rightarrow\infty$, we get  
\begin{eqnarray}
\label{eq:38.4i}
\lim_{\gamma_{\rm i}\rightarrow\infty}\left[\gamma_{\rm i} z_\rho-z_0\right]&=&\left[\frac{\rho^2}{2z_0}+\frac{z_0}{2}\right]-z_0\nonumber\\
&=&\frac{1}{2z_0}\left[{\rho^2}-{z_0^2}\right]\,.
\end{eqnarray}

Substituting Eq.~(\ref{eq:38.4i}) in Eq.~(\ref{eq:a38.4f}) we have
\begin{eqnarray}
\label{eq:88.4j}
E_z&=&  \lim_{\gamma_{\rm i}\rightarrow\infty}\left[\frac{-2ez_0}{\gamma_{\rm i}[z_0^{2}+\rho^2]}\right]\delta(z)=0\,,\nonumber \\
E_\rho&=& \frac{e}{\rho}\left[\frac{{\rho^2}-{z_0^2}}{z_0^{2}+\rho^2}+1\right]\delta(z)=\frac{2e\rho}{z_0^{2}+\rho^2}\delta(z)\,,
\end{eqnarray}
Eq.~(\ref{eq:88.4j}) is the expression for $\delta$-field (Eq.~(\ref{eq:88.4a})) that has been conjectured \cite{5,10,Ly08,ER00,FR14}, and derived by others \cite{CR15,JF15} 
using differing methods.

Of course, there is an accompanying magnetic field, ${\bf B}= {\bf v}_{\rm i}\times {\bf E}/c$, which in the limit $|v_{\rm i}|\rightarrow c$, is 
\begin{eqnarray}
\label{eq:38.4k}
B_{\phi}  =  \frac{-2e \rho}{z_0^{2}+\rho^{2}} \delta(z)\;.
\end{eqnarray}

\section{Total energy in electromagnetic field including that in $\delta$-field at time $t=0$}
It is possible to calculate analytically the electromagnetic field energy in the case of a charge moving with a uniform velocity or even of a charge moving with a uniform acceleration \cite{17,18}. We can, accordingly, calculate exactly the total energy in the field, including that in $\delta$-field, at time $t=0$. The field in the spherical region of radius $r=c\tau$ around $z_{\rm i}$, are of a charge moving with a constant proper acceleration $a$, while outside that region the field is that of a charge moving with a uniform velocity $v_{\rm i}$ (with a corresponding Lorentz factor $\gamma_{\rm i}$), the latter field turning into $\delta$-field in the limit  $\tau \rightarrow\infty$.

The electromagnetic field energy is given by the volume integral
\begin{eqnarray}
\label{eq:88.4k}
{\cal E}=\int \frac{E^{2}+B^{2}}{8\pi}\:{\rm d}V \;,
\end{eqnarray}
where ${\rm d}V= r^2 {\rm d} r (1-v \cos\theta/c)\:{\rm d}\Omega$ is the volume element (Fig.~4). 

Using the integral
\begin{eqnarray}
\label{eq:88.4l}
\int_{\Omega}\frac{(1-v \cos\theta/c)}{\gamma ^{4}(1-v \cos\theta/c)^{4}}\:{\rm d}\Omega&=&
\int_{0}^{\pi}\frac{2\pi\sin \theta}{\gamma ^{4}(1-v \cos\theta/c)^{3}}\:{\rm d}\theta\nonumber\\
&=&4\pi\;,
\end{eqnarray}
the total field energy in the volume up to $r=c\tau$, from Eqs.~(\ref{eq:38ab}) in the instantaneous rest frame at time $t=0$, is found to be
\begin{eqnarray}
\label{eq:88.4m}
{\cal E}_1 =\int_{\epsilon}^{c\tau}\frac{e^{2}}{2r^{2}}\:dr =\frac{e^{2}}{2\epsilon}-
\frac{e^{2}}{2c\tau} \;.
\end{eqnarray}
Since the integral diverges for $r\!\rightarrow \!0$, we have restricted the lower limit of $r$
to $\epsilon$, which may represent the radius of the charged particle. 
For $\tau\rightarrow \infty$, the energy in the field of an accelerated charge, at time $t=0$ (when it is instantly stationary), becomes
\begin{eqnarray}
\label{eq:88.4n}
{\cal E}_1 =\frac{e^{2}}{2\epsilon}\;.
\end{eqnarray}

This exactly is the expression for the field energy of a charge permanently at rest in an inertial frame, while in our calculations we included the contribution of the acceleration field as well, for all $r$. We find no signs of the radiated energy from the charge during its uniform accelerated motion from time $t=-\tau$ to $t=0$, as would be expected from Larmor's formula for the radiated power.
\begin{figure}[t]
\begin{center}
\includegraphics[width=7.5cm]{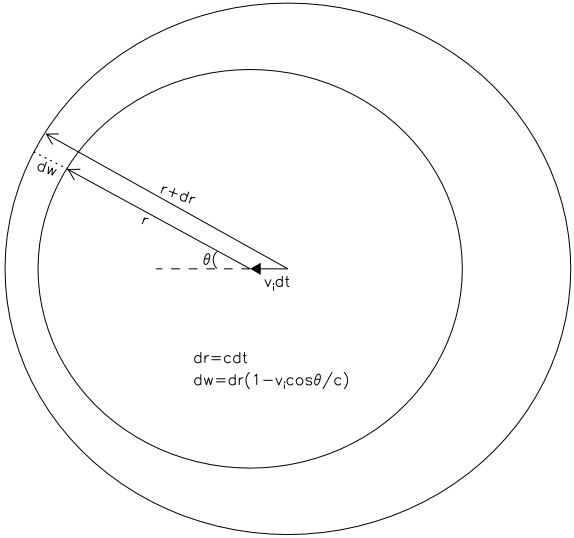}
\caption{Volume element for the computation of the field energy. Thickness of the spherical shell, shown by the dotted line, is ${\rm d}w={\rm d} r (1-v \cos\theta/c)$, along angle $\theta$, implying the corresponding volume element of the spherical shell as ${\rm d}V= r^2 {\rm d} w \:{\rm d}\Omega=r^2 {\rm d} r (1-v \cos\theta/c)\:{\rm d}\Omega$,}
\end{center}
\end{figure}

Using the integral 
\begin{eqnarray}
\label{eq:88.4o}
\int_{\Omega}\frac{(1-v \cos\theta/c)\sin ^{2}\theta}{\gamma ^{4}(1-v \cos\theta/c)^{6}}\:{\rm d}\Omega\!
&=&\!\!\int_{0}^{\pi}\frac{2\pi\sin ^{3}\theta}{\gamma_{\rm i}^{4}
(1-v_{\rm i}/c\cos \theta)^{5}}\:{\rm d}\theta\nonumber\\
&=&\frac{8\pi\gamma ^{2}}{3}\;,
\end{eqnarray}
from Eqs.~(\ref{eq:38.5b}) and (\ref{eq:38.5c}), at time $t=0$, the energy in electromagnetic field in the region $r>c\tau$ is found to be
\begin{eqnarray}
\label{eq:88.4p}
{\cal E}_2 =\frac{e^{2}}{2}\left[1+\frac{4}{3}((\gamma_{\rm i} v_{\rm i}/c )^{2}
\right]\int_{c\tau}^{\infty}\frac{{\rm d} r}{r^{2}}\;,
\end{eqnarray}
Using $(\gamma_{\rm i} v_{\rm i})^2= (a \tau)^2$ we can write 
\begin{eqnarray}
\label{eq:88.4q}
{\cal E}_2 =\frac{e^{2}}{2 c\tau}+\frac{2e^{2}a^{2}\tau}{3 c^{3}}\;.
\end{eqnarray}
For $\tau\rightarrow \infty$, the field becomes $\delta$-field, and the energy in that is
\begin{eqnarray}
\label{eq:88.4r}
{\cal E} =\frac{2e^{2}a^{2}\tau}{3 c^{3}}\;.
\end{eqnarray}
which is infinite.

It may seem, at a casual look, that this might be the elusive radiated energy, emitted by the uniformly accelerated charge during time $\tau$, at the rate, $2e^{2}a^{2}/3c^{3}$,  given by Larmor's radiation formula. In fact, it has been claimed that the radiation emitted from the uniformly accelerated charge goes beyond the horizon, the regions of space-time inaccessible to an observer co-accelerating with charge \cite{10,AL06}. Now, 
the only electromagnetic power that ever goes beyond the horizon at $z=0$ is that in the $\delta$-field. 
However, the $\delta$-field, in reality, has no causal relation with the charge during its uniform acceleration. All fields, originating from the accelerating charge positions, lie in the region $z>0$ at time $t=0$ and the radiation, if any, from the accelerating charge should also lie only there. 
In fact, because of a rate of change of
acceleration $\dot{\bf a}$, at time $t=-\tau$, an event with which the $\delta$-field has a causal relation, the charge moving with velocity $ v_{\rm i}$  undergoes radiation losses, at a rate $\propto -\gamma_{\rm i} {\bf v}_{\rm i}\cdot\dot{\bf a}$ \cite{68a},
owing to the Abraham-Lorentz radiation reaction formula \cite{1,2,abr05,16,68b}, thereby with a total energy loss 
\begin{eqnarray}
\label{eq:88.4s}
\Delta {\cal E} &=&\frac{-2e^2}{3c^{3}}\int_{-\tau_{-0}}^{-\tau_{+0}}\gamma_{\rm i} {\bf v}_{\rm i}\cdot\dot{\bf a}\:{\rm d}t=\frac{-2e^2}{3c^{3}}\gamma_{\rm i} {\bf v}_{\rm i}\cdot\int_{-\tau_{-0}}^{-\tau_{+0}}\dot{\bf a}\:{\rm d}t\nonumber\\
&=&\frac{2e^2}{3c^{3}}\gamma_{\rm i} |v_{\rm i}\:a|=\frac{2e^2a^{2}\tau}{3c^{3}}\,.
\end{eqnarray}
which exactly is the electromagnetic field energy in the $\delta$-field (Eq.~(\ref{eq:88.4r})).
\section*{Acknowledgements} 
I thank Jerrold Franklin for his comments and suggestions on the manuscript.
 
\section*{Appendix}
\appendix
\section{Electromagnetic field of a uniformly accelerated charge in terms of its ``real-time'' motion}
We want to express the electromagnetic field at some event (location and instant of time) with respect to the motion of the charge also for the same instant. It may not be possible to do so in a general case where the acceleration of the charge may be changing with time. However, for a uniformly  accelerated charge, it is possible to solve the expressions for electromagnetic field, not necessarily in terms of motion of the charge at the retarded time, instead wholly in terms of the ``real-time'' motion of the charge \cite{5,10}. 

Let $P(\rho,z$) be the field point where we want to determine the electromagnetic field of a uniformly accelerated charge at some instant of time $t$, in terms of the ``real-time'' position of the charge $z_{\rm e}$ and its velocity $v_0={a}t/\gamma_{0}$, implying 
\begin{eqnarray}
\label{eq:a88}
\gamma_{0}v_0/c=ct/z_0
\end{eqnarray}
with $\gamma_0 =[{1+(ct/z_0)^2}]^{1/2}$. 
From Eq.~(\ref{eq:38ad}), we have $z_{\rm e}=[{z_0^2+c^2t^2}]^{1/2}$. Thus all we require is the quantity $r\gamma_{\rm r}(1-v_{\rm r} \cos\theta/c)$ in Eqs.~(\ref{eq:38aaa}) and (\ref{eq:38ba}) to be expressed in terms of $\rho$, $z$, $z_0$ and $t$. 

Let $z_{\rm r}$ be the position of the charge at the corresponding retarded time $t_{\rm r}=t-r/c$, where it had the velocity $v_{\rm r} = at_{\rm r}/\gamma_{\rm r}$. From Eq.~(\ref{eq:38ad}), we also have $z_{\rm r}=[{z_0^2+c^2t_{\rm r}^2}]^{1/2}$.
\begin{figure*}[t]
\begin{center}
\includegraphics[width=\linewidth]{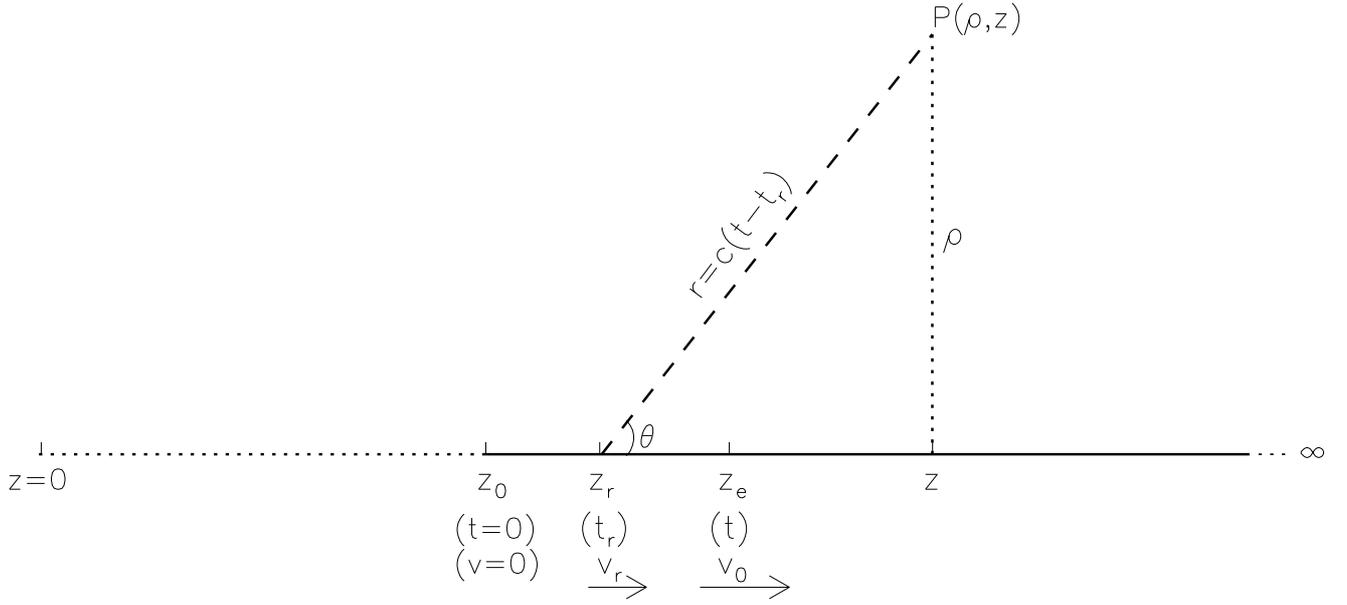}
\caption{A charge, moving with a uniform proper acceleration, ${\bf a} =\gamma^{3} \dot{\bf v}$, along the z-axis, and coming from $z=\infty$ at $t=-\infty$ with an initial velocity along  $-z$ direction, is decelerated and comes to rest momentarily at $z=z_0$ at time  $t=0$. The charge afterwards moves along the z-axis with an ever increasing velocity due to its constant proper acceleration. We want to determine  electromagnetic field at time $t$, at a point $P (\rho,z)$, which is at a distance $r=c(t-t_{\rm r})$ from the charge position $z_{\rm r}$ at the corresponding retarded time $t_{\rm r}$, when it had a velocity $v_{\rm r}$. The charge at time $t$ is at a location $z_{\rm e}$, moving with a ``present'' velocity $v_0$.}
\end{center}
\end{figure*}
Then, we have
\begin{eqnarray}
\label{eq:a38}
z_{\rm r}^2=z_0^2+c^2t_{\rm r}^2
={z_0^2+c^2t^2}+r^2-2rct\,.
\end{eqnarray}
Now, from Fig. 5, we have the relations, $\nonumber z_{\rm r}=z-r\cos\theta, \, \rho=r\sin\theta$, which gives us
\begin{eqnarray}
\label{eq:a38a}
z_{\rm r}^2+\rho^2=z^2+r^2-2 r z \cos \theta\,.
\end{eqnarray}
From Eqs.~(\ref{eq:a38}) and  Eq.~(\ref{eq:a38a}), we eliminate $z_{\rm r}$, to get
\begin{eqnarray}
\label{eq:a38a1}
2rct-2rz\cos \theta={z_0^2+c^2t^2}+\rho^{2}-z^{2}\,.
\end{eqnarray}
Squaring, we get
\begin{eqnarray}
\label{eq:a38a2}
4r^2(z^2+c^2t^2-2zct\cos \theta)\;\;\;\;\;\;\;\;\;\;\;\;\;\;\;\;\;\;\;\;\;\;\;\;\nonumber\\
=({z_0^2+c^2t^2}+\rho^{2}-z^{2})^2+4z^{2}\rho^{2}\,.
\end{eqnarray}
Using Eq.~(\ref{eq:a38a1}), to eliminate the $\cos\theta$ term, we can write
\begin{eqnarray}
\label{eq:a38.a1}
4(z^2-c^2t^2)r^2+4rct(z_0^{2}+c^2t^2+\rho^{2}-z^{2})\;\;\;\;\;\;\;\;\;\;\;\;\nonumber\\
-[({z_0^2+c^2t^2}+\rho^{2}-z^{2})^{2}+4z^{2}\rho^{2}]=0\,.
\end{eqnarray}
This quadratic equation in $r$ has a solution
\begin{eqnarray}
\label{eq:a38.b1}
r=\frac{1}{2(z^2-c^2t^2)}\left[-ct({z_0^2+c^2t^2}+\rho^{2}-z^{2})\:\right.\;\;\;\;\;\;\;\;\;\;\;\;\;\;\;\;\;\;\;\nonumber\\
+\left.z[({z_0^2+c^2t^2}-\rho^{2}-z^{2})^{2}+4z_0^{2}\rho^{2}]^{1/2}\right]\,.
\end{eqnarray}
The other possible solution of the quadratic equation is ruled out from the requirement that $r>0$.\\
Also 
\begin{eqnarray}
\label{eq:a38b2}
\gamma_{\rm r}(1-v_{\rm r} \cos\theta/c)&=&\frac{z_{\rm r}}{z_0}-\frac{at_{\rm r}\cos\theta}{c}\nonumber\\
&=&\frac{z-r\cos\theta}{z_0}-\frac{c(t-r/c)\cos\theta}{z_0}\nonumber\\
&=&\frac{z-ct\cos\theta}{z_0}\,.
\end{eqnarray}
From which we get
\begin{eqnarray}
\label{eq:a38b3}
[r\gamma_{\rm r}(1-v_{\rm r} \cos\theta/c)]^2=\frac{(rz-rct\cos\theta)^2}{z_0^2}\nonumber\\
\;\;\;\;\;\;\;\;\;\;\;\;\;=\frac{r^2(z^2+c^2t^2-2zct\cos \theta)-c^2t^2\rho^{2}}{z_0^2}\,.
\end{eqnarray}
Substituting Eq.~(\ref{eq:a38a2}) in Eq.~(\ref{eq:a38b3}), and after some simplification we get
\begin{eqnarray}
\label{eq:a38a4}
r\gamma_{\rm r}(1-v_{\rm r} \cos\theta/c)&=&\frac{[({z_0^2+c^2t^2}-\rho^{2}-z^{2})^{2}+4z_0^{2}\rho^{2}]^{1/2}}{2z_0}\nonumber\\
&=&\frac{\xi}{2z_0}\,,
\end{eqnarray}
where 
\begin{eqnarray}
\label{eq:a38ab2}
\xi=[({z_0^2+c^2t^2}-\rho^{2}-z^{2})^{2}+4z_0^{2}\rho^{2}]^{1/2}.
\end{eqnarray}

From Eqs.~(\ref{eq:a38a4}) and (\ref{eq:a38b2}), we can also write
\begin{eqnarray}
\label{eq:a38a5}
r=\frac{\xi}{2(z-ct\cos\theta)}\,.
\end{eqnarray}

Substituting Eqs.~(\ref{eq:a38a4}) in  Eqs.~(\ref{eq:38aaa}) and (\ref{eq:38ba}), and using Eqs.~(\ref{eq:a88}), (\ref{eq:a38a1}) and (\ref{eq:a38a5}), we get components of the electromagnetic field at the field point $P(\rho,z)$ at time $t$, as 
\begin{eqnarray}
\label{eq:a38ab1}
E_{z}&=& \frac{4e z_0^{2}r\cos\theta}{r\xi^{2}}- \frac{8e z_0^{3} r\sin^2\theta\gamma_0{v}_0}{c\xi^{3}}\nonumber\\
&=&4ez_0^{2}\frac{2r\cos\theta(z-ct\cos\theta)-2r ct\sin^2\theta}{\xi^{3}}\nonumber\\
&=&-4ez_0^{2}({z_0^2+c^2t^2}+\rho^{2}-z^{2})/\xi^{3}\nonumber\\
E_{\rho}&=&\frac{4e\rho z_0^{2}}{r\xi^{2}}+\frac{8e\rho z_0^{3}\gamma_0{v}_0\cos\theta}{c\xi^{3}}\nonumber\\
&=&8ez_0^{2}\rho\frac{(z-ct\cos\theta)+ct\cos\theta}{\xi^{3}}=8ez_0^{2}\rho z/\xi^{3}\nonumber\\
B_{\phi}&=&\frac{e\rho \gamma_0{v}_0/c}{\gamma ^{3}r^{3}(1- v_{\rm r}\cos\theta/c)^{3}}=8ez_0^{2}\rho ct/\xi^{3}\;,
\end{eqnarray}
the remaining  field components are zero.
 
Of course these more general expressions (Eq.~(\ref{eq:a38ab1})), along with (\ref{eq:a38ab2})), reduce to the simpler ones derived for the instantaneous rest frame (Eq.~(\ref{eq:38ab1})), along with (\ref{eq:38ab2})), by putting $t=0$.
 

\end{document}